# Spatially-Resolved Atmospheric Turbulence Sensing with Two-Dimensional Orbital Angular Momentum Spectroscopy

Wenjie Jiang[1,†], Mingjian Cheng[1,†], Lixin Guo[1,*], and Andrew Forbes[1,2,*]

[1] School of Physics, Xidian University, South Taibai Road 2, Xi'an 710071 Shannxi, China
[2] School of Physics, University of the Witwatersrand, Private Bag 3, Johannesburg 2050, South Africa
†These authors contributed equally
*These authors jointly supervised this work

Atmospheric turbulence characterization is crucial for technologies like free-space optical communications. Existing methods using a spatially-integrated one-dimensional (1D) orbital angular momentum (OAM) spectrum, $P(m)$, obscure the heterogeneous nature of atmospheric distortions. This study introduces a two-dimensional (2D) OAM spectroscopy, $P(m, n)$, which resolves the OAM spectrum (topological charge $m$) across discrete radial annuli (index $n$). Integrating this high-dimensional spectral analysis with a Support Vector Machine (SVM) classifier significantly improves the accuracy of atmospheric turbulence parameter inversion. The full potential of complex probe beams, such as multi-ringed Bessel-Gaussian beams, is realized with this radially-resolved 2D analysis. Through a co-design of the probe beam's spatial structure and the OAM spectral analysis dimensionality, a median classification accuracy of 85.47% was achieved across 20 turbulence conditions, a 23% absolute improvement over 1D techniques. The radial index also mitigates insufficient OAM spectral range, and a targeted feature-selection protocol addresses noise from low signal-to-noise ratio outer radial regions. This framework emphasizes co-design of the optical probe field and its OAM spectral analysis for enhanced fidelity in turbulence characterization.

## Introduction

Real-time, high-fidelity characterization of atmospheric turbulence is an enabling technology for next-generation optical systems. Applications spanning gigabit-per-second free-space optical (FSO) communications[1-2], precise remote sensing[3], and diffraction-limited astronomical imaging[4-5] are fundamentally constrained by stochastic fluctuations in the atmospheric refractive index. These fluctuations induce phase distortions in propagating wavefronts, leading to deleterious effects like signal fading, beam wander, and scintillation that degrade system performance and reliability[6]. While adaptive optics (AO) systems can compensate for these aberrations, their efficacy is directly governed by the accuracy and latency of the underlying turbulence sensing mechanism[7]. Traditional sensors, such as scintillometers or Shack-Hartmann wavefront sensors, often yield spatially-averaged or low-resolution measurements, providing an incomplete picture of the channel that may be insufficient for future systems demanding detailed, real-time channel state information[8].

Structured light fields, particularly vortex beams carrying orbital angular momentum (OAM), have emerged as powerful tools for atmospheric probing[9]. A vortex beam with topological charge $m$ features a helical phase front, $\exp(im\varphi)$, and an $m$-fold rotational symmetry[10]. When such a beam traverses a turbulent medium, atmospheric phase perturbations induce coupling between OAM modes, scattering optical power from the initial mode $m$ to adjacent modes $m+\Delta m$[11]. This phenomenon has become central to a recent paradigm shift towards data-driven atmospheric sensing, where machine learning is attracting widespread attention for turbulence characterization[12-13]. While early machine learning approaches successfully utilized the distorted intensity profiles (i.e., speckle patterns) of propagated probe beams as training data, the OAM spectrum is now recognized as a sensitive and structured dataset, often better suited for this task[14-16]. Because the morphology of the broadened OAM spectrum is intrinsically linked to the statistical properties of the turbulence, it serves as an effective fingerprint of the atmospheric channel for training robust inference models[17-18].

However, the prevailing OAM-based sensing paradigm suffers from a crucial limitation, it predominantly relies on a one-dimensional (1D) OAM spectrum, $P(m)$[12-13]. This spectrum is derived by spatially integrating the received field over the beam's entire transverse plane, thereby collapsing the rich, spatially-dependent information of the beam-turbulence interaction into a single, averaged vector[19-20]. This averaging process renders the 1D spectrum insensitive to the radial dependence of turbulence-induced distortions, representing a bottleneck in characterizing the heterogeneous, multi-scale nature of atmospheric turbulence. To overcome this constraint, we introduce and formalize the concept of a radially-resolved two-dimensional (2D) OAM spectrum, $P(m, n)$. Instead of a single spectrum $P(m)$, we compute a matrix $P(m, n)$, which represents the OAM spectrum as a function of both topological charge $m$ and a discrete radial annulus index $n$. By partitioning the beam's transverse plane into concentric annuli, this technique maps the OAM modal distribution radially, preserving the spatial information of the phase perturbations and generating a high-dimensional fingerprint of the beam-turbulence interaction. The resulting 2D spectrum offers a more complete physical picture, revealing how modal crosstalk and energy degradation manifest across the beam's transverse profile. Such high-dimensional data are exceptionally well-suited for analysis via machine learning, which excels at identifying complex features in large datasets for atmospheric parameter inversion[12]. We synergize this sensing modality with the Support Vector Machine (SVM)[21], an algorithm renowned for its effectiveness in classification problems with high-dimensional feature spaces[22-23], to learn the intricate mapping between features of the 2D OAM spectrum and the defining parameters of the atmospheric turbulence.

In this work, we demonstrate that the synthesis of structured probe beams, 2D OAM spectroscopy, and machine learning establishes a highly accurate and robust methodology for atmospheric turbulence sensing. We show that by co-designing the probe beam's structure and the dimensionality of the spectral analysis, we achieve a substantial enhancement in the accuracy of turbulence parameter inversion. This work introduces a superior methodology for environmental optics and provides new physical insights into the spatially-dependent nature of light-turbulence interactions, paving the way for more intelligent and resilient optical systems.

## Theoretical Model
### Probe Beam Generation and Structure

To resolve the radial structure of atmospheric turbulence, the ideal optical probe must possess a well-defined and controllable radial scale. While both Laguerre-Gaussian (LG) and Bessel-Gaussian (BG) beams carry OAM and feature annular intensity profiles, BG beams are uniquely suited for this application. A BG beam is characterized by a single, dominant transverse wave vector, $k_r$, which imparts a well-defined radial frequency across the beam's profile[24-25]. This contrasts with LG beams, whose multiple concentric rings arise from the zeros of a polynomial and exhibit more complex evolution during propagation. The BG beam's single characteristic spatial frequency allows its transverse profile to be naturally decomposed into equidistant concentric annuli, enabling a more precise isolation of turbulence-induced phase perturbations within specific radial zones.

The complex amplitude of a BG beam at the source plane ($z=0$) in cylindrical coordinates ($r$, $\varphi$) is given by[26]:

$$U(r,\phi) = A_0 J_m(k_r r) \exp(im\phi) \exp\left(-\frac{r^2}{\omega_0^2}\right) \quad (1)$$

where $A_0$ is an amplitude constant, $J_m(x)$ is the $m$-th order Bessel function of the first kind, $m$ is the topological charge, and $w_0$ is the Gaussian waist that apodizes the otherwise infinite-extent Bessel function. The key parameter is the transverse wave number $k_r = k\sin\theta_0$, where $k$ is the optical wave number and $\theta_0$ is the conical half-angle of the plane waves forming the beam. This parameter $\theta_0$ provides direct, continuous control over the radial scale of the beam's intensity rings, as shown in Fig. 1. Increasing $\theta_0$ increases $k_r$, leading to a higher density of rings. Figure 1a demonstrates this adaptable radial structure for a BG beam, contrasting it with the discrete ring structure of an LG beam (Fig. 1b), where the number of rings is tied to the integer radial index $p$. The continuous control over the BG beam's fundamental radial scale via $\theta_0$ makes it a superior and more versatile tool for our annular decomposition analysis.

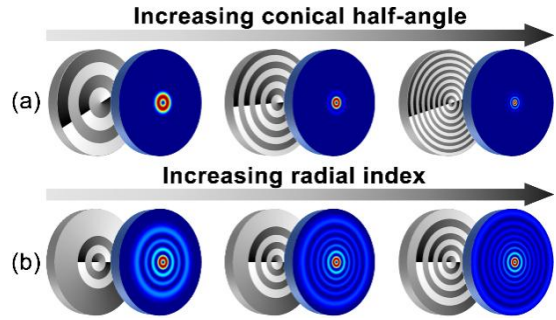

**Fig. 1 Comparison of radial intensity and phase control for Bessel-Gaussian and Laguerre-Gaussian beams.**
**a**, Intensity (right) and phase (left) profiles of Bessel-Gaussian (BG) beams demonstrating continuous control over ring density via the conical half-angle $\theta_0$. **b**, Intensity (right) and phase (left) profiles of Laguerre-Gaussian (LG) beams, where the discrete radial index $p$ dictates the number of concentric rings.

**Statistical Description of Atmospheric Turbulence**

Atmospheric turbulence is modeled as a random field of spatiotemporal refractive index fluctuations. The statistical properties of these fluctuations are well-approximated by the Kolmogorov theory, which is parameterized by metrics that quantify turbulence strength and scale.

The primary measure of turbulence strength is the refractive index structure constant, $C_n^2$. For a horizontal propagation path of length $L$, the path-integrated effect of turbulence is conveniently captured by the Fried parameter, $r_0$[27]:

$$r_0 = \left(0.423 k^2 \int_0^L C_n^2(z) dz\right)^{-3/5} \quad (2)$$

where $k=2\pi/\lambda$ is the wave number with $\lambda$ being wavelength. The Fired parameter $r_0$ represents the aperture diameter over which the root-mean-square (RMS) phase distortion is approximately 1 radian. It provides a direct measure of wavefront quality, with smaller values of $r_0$ indicating stronger turbulence.

The range of turbulent eddy sizes is bounded by an inner scale, $l_0$, and an outer scale, $L_0$. To encapsulate the influence of this range, we employ the von Kármán spectrum[6] and define a dimensionless turbulence Reynolds number, $Re=(l_0/L_0)^{-4/3}$. This parameter quantifies the breadth of the inertial range of turbulence; a larger $Re$ signifies a wider spectrum of eddy scales, leading to more complex scattering and enhanced modal crosstalk[28].

The combined effect of these parameters on the probe beam is quantified by the scintillation index, $\sigma_I^2$, the normalized variance of intensity fluctuations[6]. In the weak fluctuation regime, it is related to the Rytov variance, $\sigma_R^2$:

$$\sigma_I^2 = 3.86\sigma_R^2 \left[(1+1/Q_m^2)^{11/12} \sin\left(\frac{11}{6}\tan^{-1} Q_m\right) - \frac{11}{6} Q_m^{-5/6}\right] \quad (3)$$

The Rytov variance $\sigma_R^2 = 1.23 C_n^2 k^{7/6} L^{11/6}$ serves as a robust indicator for classifying turbulence regimes. Here, we simulated 20 distinct turbulence conditions by varying the Fried parameter from $r_0 = 0.005$ m (strong) to $r_0 = 0.02$ m (moderate-to-weak) and the Reynolds number between 1,574 and 20,000, under a fixed outer scale $L_0 = 10$ m. Table 1 summarizes the corresponding scintillation indices. These two parameters, ($r_0$, $Re$), serve as the ground truth labels for training and evaluating our machine learning model. The complete dataset comprised 20000 samples, uniformly distributed across 20 turbulence conditions (approximately 1000 samples per conditions).

**Table 1.** Atmospheric Turbulence Strength (Scintillation Index, $\sigma_I^2$) for Selected Fried Parameters ($r_0$) and Reynolds Numbers (Re). The values span weak ($\sigma_I^2 < 1$), moderate ($\sigma_I^2 \approx 1$), and strong ($\sigma_I^2 > 1$) turbulence regimes.

| $\sigma_I^2$ | $r_0$=0.005m | $r_0$=0.01m | $r_0$=0.015m | $r_0$=0.02m |
|---|---|---|---|---|
| Re=1574 | 5.062 | 1.594 | 0.811 | 0.502 |
| Re=2000 | 6.166 | 1.942 | 0.988 | 0.612 |
| Re=3960 | 8.964 | 2.824 | 1.437 | 0.889 |
| Re=10000 | 11.315 | 3.564 | 1.813 | 1.123 |
| Re=20000 | 12.186 | 3.839 | 1.953 | 1.209 |

### The 2D OAM Spectrum as a Spatially-Resolved Observable

To probe the turbulent medium, we require an observable that is highly sensitive to the turbulence induced phase distortions. While the conventional 1D OAM spectrum, $P(m)$, provides such a measure, it is limited by spatial averaging. By integrating over the beam's transverse plane, it collapses all spatial information into a single dimension, rendering it incapable of resolving the heterogeneous nature of turbulence[19-20].

To overcome this, we introduce the 2D OAM spectrum, $P(m, n)$, as a more powerful observable. This spectrum resolves OAM modal power as a function of both the azimuthal mode ($m$) and a discrete radial index ($n$), which corresponds to one of several concentric annuli partitioning the beam's cross-section. By retaining this radial dimension, the 2D OAM spectrum encodes the spatial distribution of modal crosstalk, providing a rich dataset that captures how turbulence perturbs different annular regions of the probe beam. The conventional 1D spectrum is thus the degenerate case (effectively $n=1$, where all radial information is integrated out).

The power of this approach is visualized in Fig. 2, which simulates the progressive degradation of a beam's 2D OAM spectrum during propagation. Initially, the spectrum is a sharp peak at the input mode ($m=1$) (Fig. 2a). With increasing distance, the spectrum broadens along both the topological charge axis ($m$), reflecting modal crosstalk, and the radial axis ($n$), reflecting non-uniform turbulent effects (Fig. 2b-e). This radial evolution is a direct signature of how turbulence affects the beam's profile differently at its core versus its periphery, information that is completely lost in a 1D spectral analysis. The resulting 2D distribution, $P(m, n)$, serves as a high-fidelity fingerprint of the beam-turbulence interaction, forming the theoretical foundation of our enhanced sensing methodology.

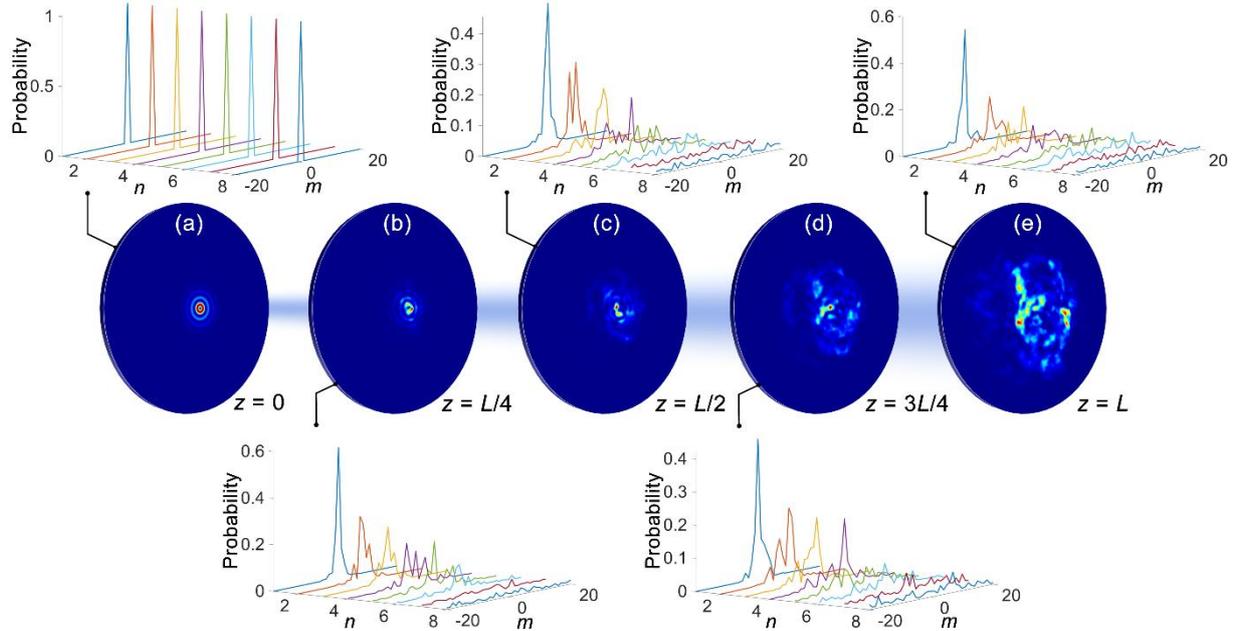

**Fig. 2 Evolution of transverse intensity and 2D OAM spectrum during turbulent propagation. a-e,** Snapshots showing the progressive degradation of the transverse intensity profile (top row) and the corresponding radially-resolved OAM spectrum, $P(m,n)$, (bottom row) as a BG beam with initial topological charge $m=1$ propagates through increasing distances ($z$) within a turbulent medium. The initial OAM spectrum (a) disperses into multiple OAM and radial modes due to turbulence-induced modal crosstalk and non-uniform phase perturbations (b-e).

### Results and Discussion

Our primary objective was to develop and optimize a robust protocol for atmospheric turbulence characterization using vortex beams and their OAM spectra. This investigation systematically explored the intricate interplay between the probe beam's spatial profile and the dimensionality of OAM spectral analysis, aiming to enhance the accuracy and reliability of turbulence parameter inversion. The efficacy of various configurations was quantified by the median classification accuracy of an SVM trained to distinguish between 20 distinct atmospheric turbulence conditions, each characterized by specific Fried parameter ($r_0$) and Reynolds number ($Re$) values. To ensure statistical robustness, all classification accuracy results presented herein are derived from 30 cross-validations on the shuffled dataset. Our analysis critically compares the performance of the conventional, spatially-integrated 1D OAM spectrum, $P(m)$, with our proposed radially-resolved 2D OAM spectrum, $P(m, n)$.

We first investigated the influence of the probe beam's radial structure on turbulence classification accuracy. We employed BG beams configured with 4, 8, and 16 concentric rings, corresponding to varying conical half-angles ($\theta_0$) (Fig. 3a-c). As shown in Fig. 3d-f, the utility of a more complex radial profile is highly dependent on the chosen OAM spectral analysis method. For the conventional 1D OAM spectrum, increasing the number of rings consistently degraded classification accuracy. This degradation occurs because while a finer radial structure facilitates more intricate interaction with turbulent phase distortions, the inherent spatial integration of $P(m)$ averages out this detailed information, resulting in a less distinct spectral signature for turbulence characterization.

In contrast, when employing the 2D OAM spectrum ($n>1$), classification accuracy scaled monotonically with the number of rings. By preserving the radial distribution of phase information, the 2D spectrum $P(m, n)$ effectively leverages the detailed beam-turbulence interaction across different radial zones. This demonstrates that a higher-dimensional OAM spectral analysis is essential to fully exploit the advantages offered by spatially complex probe fields. Furthermore, for any given beam configuration, classification accuracy rapidly saturated with increasing radial bins, typically stabilizing for $n \geq 4$. This indicates that a relatively modest radial sampling resolution is sufficient to capture the salient features of the turbulence-induced OAM spectrum, providing a clear path toward optimizing computational efficiency without sacrificing accuracy.

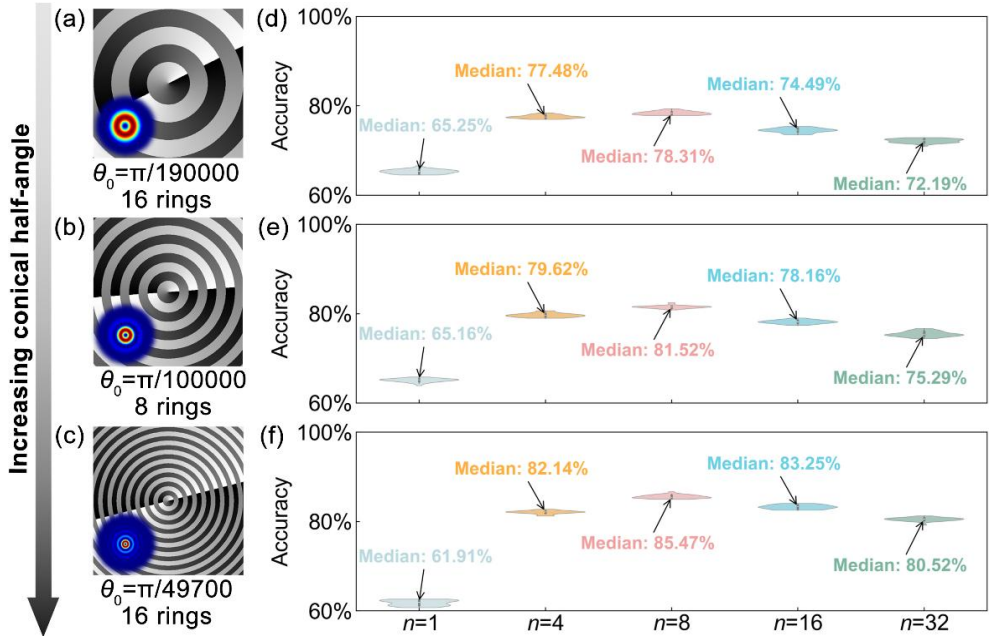

**Fig. 3 Influence of BG probe beam radial structure on Support Vector Machine (SVM) classification accuracy. a-c**, Transverse phase profiles of BG beams engineered with 4, 8, and 16 concentric rings, achieved by varying the conical half-angle $\theta_0$. **d-f**, Corresponding median SVM classification accuracy for 20 distinct turbulence conditions, plotted as a function of the number of radial divisions, $n$, employed in the OAM spectral analysis for each beam configuration. Performance of the conventional 1D OAM spectrum ($n=1$) is shown for comparison.

The influence of the initial topological charge, $m$, on classification accuracy was investigated by evaluating all OAM spectra with $m$ ranging from 1 to 5. We evaluated 16-ring BG beams with initial topological charges of $m=1$, 3, and 5 (Fig. 4a-c). Our results (Fig. 4d-f) reveal that for traditional 1D OAM spectra, classification accuracy is significantly influenced by the initial topological charge, with higher-order modes ($m>1$) generally yielding improved performance. This enhancement originates from the steeper azimuthal phase gradient inherent in high-order OAM modes, which acts as a more effective rotational 'lever'. This mechanism effectively converts phase disturbances caused by turbulence into larger and more distinct changes within the OAM spectrum. However, for 2D OAM spectrum ($n>1$), the initial topological charge has a less pronounced effect on classification accuracy. The 2D OAM spectrum places greater emphasis on the radial profile as a critical degree of freedom, leading to a diminished relative impact of the initial topological charge.

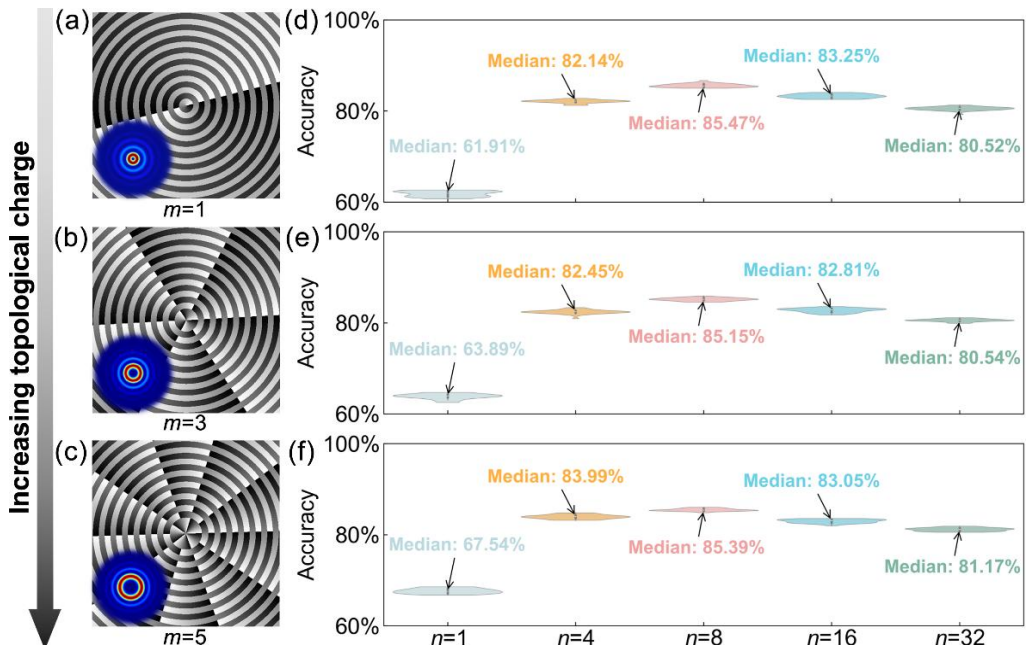

**Fig 4 Impact of initial topological charge and radial sampling resolution on turbulence classification accuracy. a-c,** Simulated intensity (left) and phase (right) profiles for 16-ring Bessel-Gaussian (BG) beams with initial topological charges of $m=1$, 3, and 5, respectively. **d-f**, Violin plots illustrating the distribution of SVM

classification accuracy as a function of the number of radial divisions, $n$, in the OAM spectral analysis for each corresponding initial topological charge.

Subsequently, to dissect the contribution of the OAM data, the effect of its spectral range on classification performance was analyzed in Fig. 5. We evaluated 16-ring BG beams with initial topological charges of $m=1$ and tested the OAM spectra with radial index $n=1$ and $n=8$. As shown in Fig. 5, increasing the spectral range from 11 modes ($m \in [-5,+5]$) up to 41 modes ($m \in [-20,+20]$) resulted in a monotonic increase in classification accuracy, confirming that a wider spectrum contains more discriminative information. Accordingly, a spectral range of $m \in [-20,+20]$ was adopted for all primary experiments to maximize the informational content available from the spectral samples. Notably, even when the spectral range was expanded only to 21 modes ($m \in [-10,+10]$), the performance of the 2D OAM spectrum (with $n=8$) already surpassed that of the 41-mode 1D OAM spectrum. This indicates that the 2D OAM spectrum, by incorporating an additional radial dimension, achieves superior performance even within a relatively narrower OAM range. The inclusion of radial variations enables the model to extract richer information from the spectral data. The practical implications of this co-design approach, where the probe's properties are optimized in tandem with the analysis method, are vividly demonstrated by the confusion matrices presented in Fig. 6. The normalized confusion matrix for the optimized configuration ($m=1$, $n=8$, $\theta_0=\pi/49700$) exhibits a pronounced diagonal (Fig. 6b), particularly under moderate to strong turbulence conditions (label≥8, corresponding to higher scintillation indices in Table 1), where classification accuracy is notably high. These results confirm that this synergistic co-design is essential for achieving high-fidelity turbulence inversion.

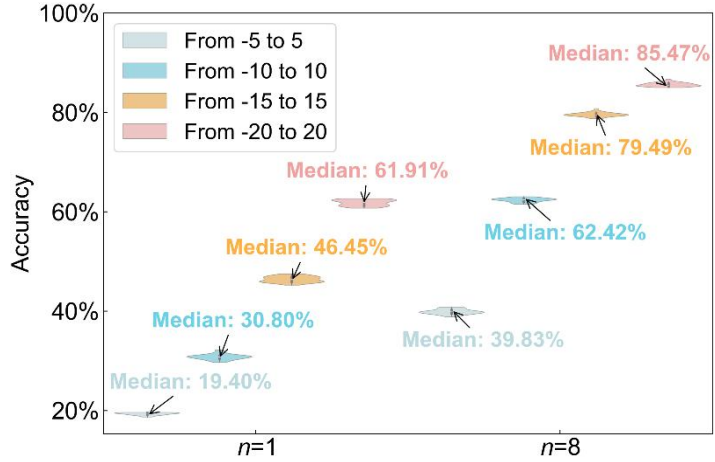

**Fig 5 Influence of OAM spectral range and dimensionality on SVM classification accuracy.** Median classification accuracy for a 16-ring BG beam with $m=1$, demonstrating the effect of expanding the OAM spectral range (number of $m$ modes) for both 1D OAM spectra ($n=1$) and 2D OAM spectra with 8 radial divisions ($n=8$). The 2D spectrum with a narrower OAM range can outperform the 1D spectrum with a wider OAM range.

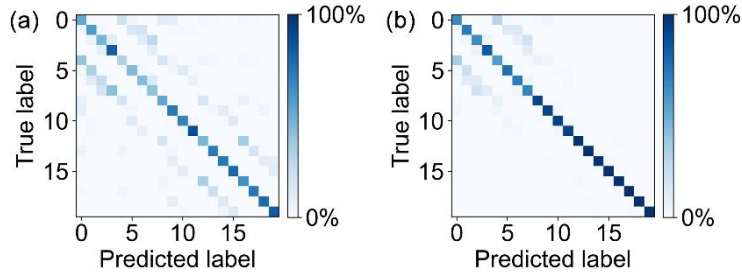

**Fig 6 Comparison of normalized confusion matrices for 1D and 2D OAM spectral analysis in turbulence classification.** Normalized confusion matrices derived from the test set, illustrating classification performance for a 16-ring BG probe beam with $m=1$ and $\theta_0=\pi/49700$. **a**, Performance using a conventional 1D OAM spectrum ($n=1$). **b**, Performance using a radially-resolved 2D OAM spectrum with $n=8$ radial divisions. Labels 1-20 correspond to the distinct turbulence conditions detailed in Table 1.

While the 2D OAM spectrum contains extensive information, not all features contribute uniformly to classification fidelity. The outermost annuli of the propagated beam typically exhibit lower signal intensity (Fig. 7a-d), rendering their information highly susceptible to noise. The indiscriminate inclusion of these noise-dominated features risks contaminating the SVM classifier's feature vector, potentially degrading model generalization and promoting overfitting to stochastic, non-informative variations. To mitigate this, we implemented a targeted feature selection strategy, systematically training an SVM classifier using truncated 2D OAM spectra, wherein only the innermost $j$ radial layers of a full n-layer spectrum were utilized.

Our analysis (Fig. 7e-g) reveals a non-monotonic relationship between classification accuracy and the number of included radial layers ($j$). Accuracy initially increases as more relevant information is incorporated from the central, higher signal-to-noise ratio regions. However, it subsequently declines with the inclusion of noise-dominated outer layers. While optimal $j$ values varied depending on specific turbulence conditions, a consistent finding was that when $j$ is approximately half of n, the system approaches its peak accuracy rate. This configuration, utilizing only the innermost half of the radial information, emerges as a recommended option for robust classification. Beyond these optimal points, performance consistently declined as the inclusion of noise-corrupted outer annuli degraded the model's predictive fidelity. This confirms that judicious feature selection, based on signal-to-noise considerations, is crucial for optimizing the sensing protocol for real-world applications, ensuring both accuracy and computational efficiency.

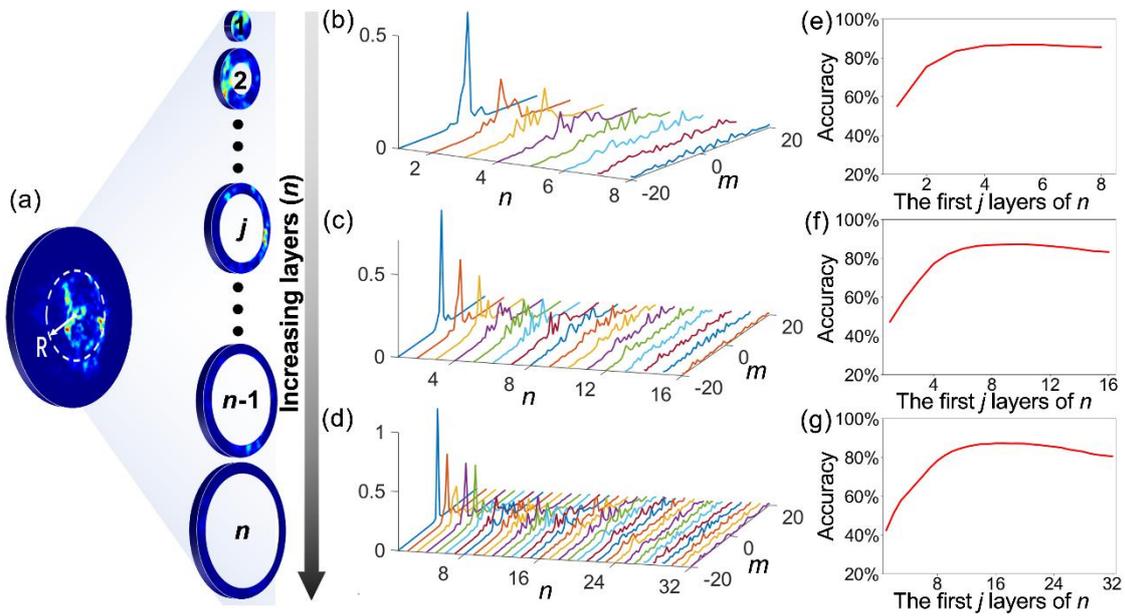

**Fig. 7 Optimization of radial sampling depth and feature selection for enhanced turbulence classification. a**, Schematic representation of a beam's transverse cross-section segmented into n concentric annuli for 2D OAM spectral analysis. **b-d**, Illustrative radially-resolved OAM spectra, $P(m,n)$, of a beam propagated through strong turbulence, computed with total radial divisions of $n$=8, 16, and 32, respectively. **e-g**, Corresponding median SVM classification accuracy as a function of the number of innermost annuli, $j$, included in the analysis (where $j<n$), demonstrating the effect of discarding noise-dominated outer radial regions.

## Conclusion

In summary, we present a novel paradigm for atmospheric turbulence characterization based on the synergistic co-design of structured optical probes and a radially-resolved 2D OAM spectral analysis. This approach significantly advances beyond conventional 1D OAM spectroscopy, which intrinsically averages out crucial spatial information. Our proposed 2D OAM spectroscopy method, coupled with an SVM classifier, substantially enhances the fidelity of turbulence parameter inversion, achieving median classification accuracies of 85.47% for optimized probe beam configurations.

Our investigation yields several critical physical insights essential for designing robust and high-performing sensing protocols. First, we demonstrate that the full potential of complex, radially-structured probe beams is unlocked only through spatially-resolved 2D OAM spectroscopy, as conventional 1D methods intrinsically average out the intricate spatial information critical for accurate inversion. Second, while higher-order topological charges can act as sensitive indicators for transducing turbulence-induced phase perturbations into spectral changes, their relative effectiveness is diminished by the introduction of an additional degree of freedom, the radial index, which captures distinct information. Third, expanding the radial index dimension can effectively mitigate the limitation of an insufficient OAM spectral range and enhance the discriminative information contained within the spectral data. Finally, we established the critical necessity of a targeted feature-selection protocol that selectively discards information from noise-corrupted outer radial regions, significantly improving the classifier's generalization and robustness by preventing overfitting to non-informative variations.

While this numerical study provides a definitive proof-of-principle, our robust framework serves as a direct blueprint for experimental implementation, validating the theoretical advantages of co-design. This methodology holds immediate applicability for enhancing the performance of FSO communication systems by furnishing high-fidelity channel state information crucial for adaptive pre-compensation or more sophisticated AO systems. Future research will explore other classes of structured beams and extend this versatile paradigm to challenging sensing environments beyond atmospheric turbulence, such as underwater optical turbulence. Ultimately, this work lays a crucial foundation for a new generation of intelligent optical sensors capable of parsing the complex, multi-scale structure of turbulent media with enhanced accuracy and insight.

## Methods

### Numerical Simulation of Atmospheric Propagation

We modeled the propagation of a vortex beam through a turbulent medium using a high-fidelity multiple phase screen (MPS) simulation based on the split-step Fourier method. This approach discretizes the continuous turbulent path of length $L$ into $N$ equally spaced, thin phase screens, separated by a distance $\Delta z = L/N$. Each screen introduces a phase perturbation that statistically emulates the cumulative effect of turbulence within that sub-interval (Fig. 8).

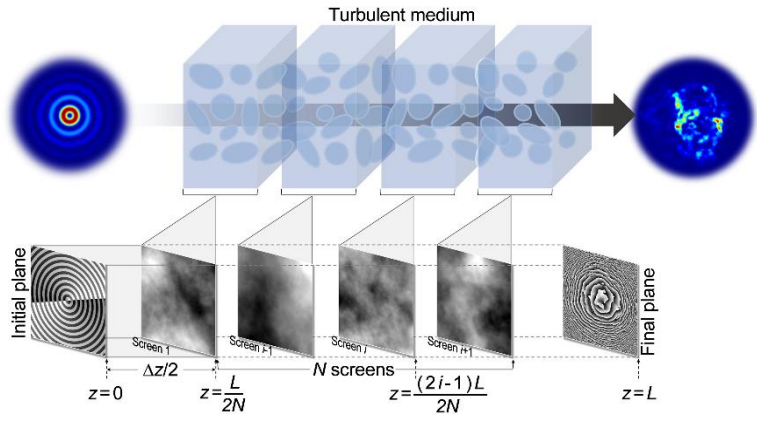

**Fig. 8 Schematic representation of the multiple phase screen (MPS) method for simulating beam propagation in the turbulence.** An initial optical field at the source plane (z=0) undergoes iterative free-space propagation steps (indicated by arrows) interspersed with discrete phase screens. Each screen applies a random phase perturbation, statistically equivalent to the cumulative effect of atmospheric refractive index fluctuations within that segment, ultimately yielding a distorted beam profile at the receiver plane (z=L).

The phase perturbations, $\phi_{turb}(x,y)$, for each screen are stochastic realizations derived from the modified von Kármán power spectral density (PSD), which accounts for both inner ($l_0$) and outer ($L_0$) scale effects[6]:

$$\Phi_n(\kappa) = 0.033 C_n^2 \frac{\exp(-\kappa^2/\kappa_m^2)}{(\kappa^2 + \kappa_0^2)^{11/6}} \quad (4)$$

where $\kappa$ is the spatial wavenumber, $\kappa_0 = 2\pi/L_0$ and $\kappa_m = 5.92/l_0$. The 2D PSD of the phase fluctuations on a single screen, $\Phi_\phi(\kappa, z)$, is related to $\Phi_n(\kappa)$ by:

$$\Phi_\phi(\kappa, z) = 2\pi k^2 \Delta z \Phi_n(\kappa) \quad (5)$$

where $\Delta z$ represents the distance between successive phase screens.

A random phase screen is generated in the frequency domain by filtering a complex Gaussian white-noise field, $a_R(k_x, k_y)$, with the phase PSD. The resulting field is then transformed to the spatial domain via an inverse Fast Fourier Transform (IFFT)[31]:

$$\phi_{turb}(x,y) = \mathrm{IFFT}\left[ a_R \sqrt{\Phi_\phi(\kappa_x, \kappa_y, z)} \right] \quad (6)$$

The beam's complex field, $U(x,y)$, is propagated iteratively. The field is first advanced through a free-space interval of length $\Delta z$ using the angular spectrum method, followed by the application of the phase perturbation from the subsequent screen:

$$U_{out}(x,y) = \mathrm{IFFT}\left\{ \mathrm{FFT}\left[ U_{in}(x,y) \exp(i\phi_{turb}(x,y)) \right] \cdot H(k_x, k_y) \right\} \quad (7)$$

where $U_{in}$ is the field distribution before passing through the phase screen, and $U_{out}$ is the field distribution after passing through the phase screen. And FFT is the Fast Fourier Transform and $H(k_x, k_y)$ is the angular spectrum transfer function:

$$H(k_x, k_y) = \exp\left[ -\frac{i\Delta z}{2k}(k_x^2 + k_y^2) \right] \quad (8)$$

This process is repeated $N$ times. The turbulence strength for each simulation is defined by the overall Fried parameter, $r_{0(total)}$, which dictates the strength of each individual screen via $r_{0(single)} = r_{0(total)} \times N$. The key simulation parameters are detailed in Table 2.

**Table 2.** Key Parameters for Numerical Propagation Simulation.

| Parameters | Value |
| --- | --- |
| Wavelength (m) $\lambda$ | 1550×10$^{-9}$ |
| Waist radius (m) $w_0$ | 0.05 |
| Screen size (m) $b$ | 0.4 |
| Resolution ratio $N_s$ | 1024*1024 |
| Distance (m) $L$ | 1000 |
| Distance between phase screens (m) $\Delta z$ | 20 |
| Mode order $m$ | 1 |
| Angular parameter $\theta_0$ | π/49700 |
| Reynolds number $Re$ | 1574, 2000, 3960, 10000 and 20000 |
| Fried parameter (m) $r_0$ | 0.005, 0.01, 0.015 and 0.02 |

## OAM Spectral Analysis

At the receiving plane (z=L), the simulated complex field $U(r, \varphi, z)$ is processed to generate the 2D OAM spectrum, P(m,n). The field's cross-section is first partitioned into n discrete, concentric annuli (Fig. 9a). For each annulus j, corresponding to a radial interval $[r_{j-1}, r_j]$, the data is isolated.

The OAM modal decomposition within each annulus is performed by projecting the corresponding field segment, $U_j(r, \varphi, z)$, onto the spiral harmonic basis, $\exp(im\varphi)$. This is efficiently computed via a Fourier transform with respect to the azimuthal coordinate, after a coordinate transformation "unwraps" the annulus into a rectangular grid in $(r, \varphi)$ space (Fig. 9b). The complex amplitude of the m-th OAM mode in the j-th annulus is given by:

$$C(m,j) = \frac{1}{\sqrt{2\pi}} \int_{r_{j-1}}^{r_j} \int_0^{2\pi} U(r,\phi,z) \exp(-im\phi) r \, d\phi \, dr \qquad (9)$$

The normalized power spectrum, $P(m, j)$, is the squared magnitude of these coefficients, normalized by the total power within that annulus:

$$P(m,j) = \frac{|C(m,j)|^2}{\sum_{l=-\infty}^{\infty} |C(l,j)|^2} \qquad (10)$$

By assembling the spectra from all n annuli, we construct the final 2D OAM spectrum, a matrix of size ($n$, $2M+1$), where $2M+1$ is the number of OAM modes analyzed (from $-M$ to $+M$). This matrix (Fig. 9c) serves as the high-dimensional fingerprint of the turbulence.

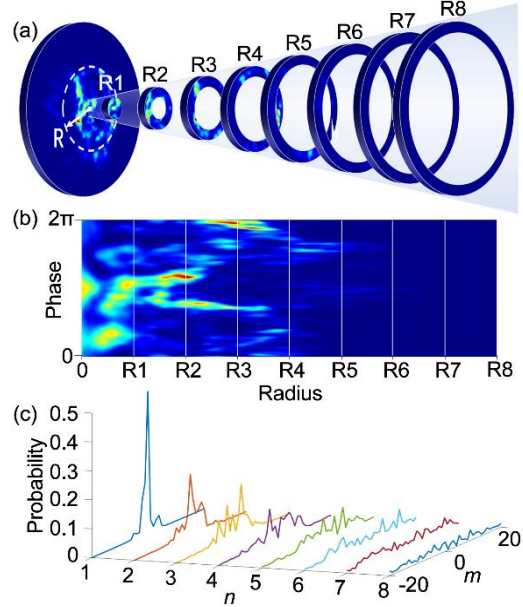

**Fig. 9 Computational framework for constructing the radially-resolved 2D OAM spectrum. a**, The complex field of the received beam is initially segmented into *n* distinct concentric annuli. **b**, Each isolated annulus undergoes a coordinate transformation from Cartesian to polar coordinates, effectively 'unwrapping' it into a rectangular grid. **c**, A Fourier transform is then applied along the azimuthal dimension of each unwrapped annulus to compute its local OAM spectrum. These individual spectra are then compiled into the 2D matrix $P(m,n)$, which represents modal power as a function of both topological charge *m* and radial index *n*.

To validate our method, we confirmed that the 1D spectrum, $P(m)$, can be perfectly recovered by coherently summing the complex amplitudes across all radial layers before calculating power: $P(m) = \frac{\sum_{j=1}^{n} |C(m,j)|^2}{\sum_{j=1}^{n} \sum_{l=-\infty}^{\infty} |C(l,j)|^2}$.

As shown in Fig. 10, the 1D spectra recovered from our 2D data for various radial divisions ($n$=4, 8, 16, 32) perfectly match the spectrum calculated directly from the full beam, confirming the self-consistency of our 2D decomposition.

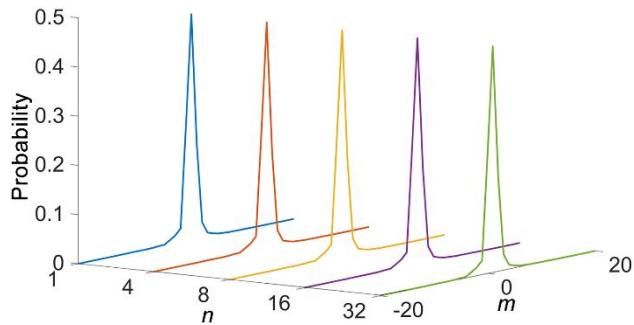

**Fig. 10 Validation of the 2D OAM spectrum construction through 1D spectrum recovery.** The conventional 1D OAM spectrum (dark blue line), calculated directly from the full beam after propagation through strong turbulence ($r_0$=5 mm, $Re$=20000), serves as a reference. Overlaid are the 1D OAM spectra (colored lines) reconstructed by summing the corresponding 2D spectrum, $P(m,n)$, across its radial index n for various total radial divisions ($n$=4, 8, 16, and 32). The perfect alignment demonstrates the self-consistency and accuracy of the 2D decomposition method.

## Machine Learning Protocol

A Support Vector Machine (SVM) was employed for the multi-class classification of the 20 distinct turbulence conditions. We used a C-support vector classification model with a radial basis function (RBF) kernel, which is effective for capturing non-linear relationships in high-dimensional feature spaces[21]. The kernel function is defined as:

$$K(x_i, y_j) = \exp\left(-\gamma \|x_i - y_j\|^2\right) \qquad (11)$$

where $x_i$ and $x_j$ are feature vectors and $\gamma$ is the kernel coefficient.

The input to the SVM was a feature vector constructed by flattening the 2D OAM spectrum matrix P(m, n) of size (*n*, 41) into a single vector of length n×41. The full dataset of simulated beam propagations was randomly split into a training set (70%) and a testing set (30%). The SVM's hyperparameters, the regularization parameter *C* and the kernel coefficient *γ*, were optimized via a grid search with 5-fold cross-validation on the training dataset to prevent overfitting and maximize generalization performance. To account for variations in data dimensionality, Fig. 11 presents our grid searches on datasets comprising OAM spectra with different radial index *n*. Figure 11a demonstrates the effect of the kernel coefficient γ on classification accuracy under a fixed regularization parameter *C*=2000. The results indicate that datasets with higher n necessitate smaller values of *γ* to achieve optimal performance in the SVM models, with the best-performing curve (*n*=8) attaining peak accuracy at $\gamma=10^{-4}$. Regarding the regularization coefficient *C*, although larger values may yield marginal improvements in accuracy, they concomitantly lead to increased computational costs and a higher risk of overfitting. As depicted in Figure 11b, accuracy improvements beyond *C*=2000 are marginal or even slightly reduced in certain cases (e.g., *n*=16 and 32), rendering further increases impractical. Thus, the optimal values were identified as *C*=2000 and $\gamma=10^{-4}$, which achieve satisfactory recognition accuracy without introducing excessive computational burden or potential overfitting. To ensure robust performance across all turbulence regimes, class weights were adjusted during training to counteract any imbalances in the dataset. The model's final performance was evaluated on the unseen testing set, with classification accuracy serving as the primary metric. For the feature selection analysis, the SVM was retrained using truncated feature vectors corresponding to specific subsets of the radial layers.

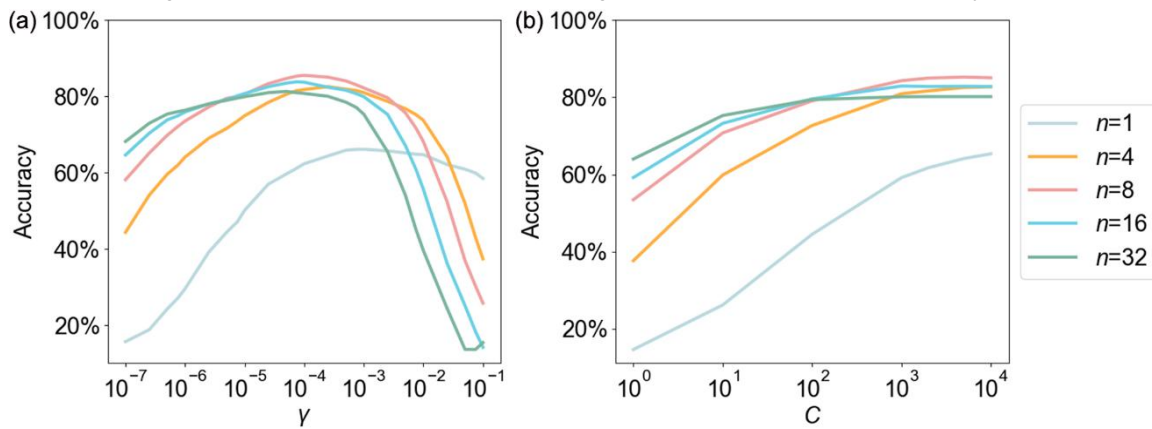

**Fig. 11 Hyperparameter optimization for SVM classification with radially-resolved OAM spectra**. **a**, Grid search results showing the impact of the kernel coefficient *γ* on classification accuracy, evaluated across datasets with varying radial division numbers (*n*) at a fixed regularization parameter *C*=2000. **b**, Grid search results illustrating the effect of the regularization parameter *C* on classification accuracy for different *n*, with a fixed kernel coefficient $\gamma=10^{-4}$.

**References**
1. Khalighi, M. A. & Uysal, M. Survey on free space optical communication: A communication theory perspective. *IEEE Commun. Surv. Tutor.* **16**, 2231–2258 (2014).
2. Zhu, X. & Kahn, J. M. Free-space optical communication through atmospheric turbulence channels. *IEEE Trans. Commun.* **50**, 1293–1300 (2002).
3. Cheng, M. et al. Metrology with a twist: probing and sensing with vortex light. *Light: Science & Applications* **14**, 4 (2025).
4. Hardy, J. W. *Adaptive optics for astronomical telescopes*. Oxford University Press (1998).
5. Roggemann, M. C. & Welsh, B. M. *Imaging through turbulence*. CRC press (2018).
6. Andrews, L. C. & Beason, M. K. *Laser beam propagation in random media: new and advanced topics* (2023).
7. Tyson, R. K. *Principles of Adaptive Optics* (CRC press, 2010).
8. Platt, B. C. & Shack, R. History and Principles of Shack-Hartmann Wavefront Sensing. *J. Refract. Surg.* **17**, S573–S577 (2001).
9. Forbes, A., de Oliveira, M. & Dennis, M. R. Structured light. *Nat. Photon.* **15**, 253–262 (2021).
10. Allen, L., Beijersbergen, M. W., Spreeuw, R. J. C. & Woerdman, J. P. Orbital angular momentum of light and the transformation of Laguerre-Gaussian laser modes. *Phys. Rev. A* **45**, 8185 (1992).
11. Paterson, C. Atmospheric turbulence and orbital angular momentum of single photons for optical communication. *Phys. Rev. Lett.* **94**, 153901 (2005).
12. Zhou, H. et al. Atmospheric turbulence strength distribution along a propagation path probed by longitudinally structured optical beams. *Nat. Commun.* **14**, 4701 (2023).
13. Chen, Z. et al. Weather sensing with structured light. *Commun. Phys.* **8**, 105 (2025).
14. Guo, Y. et al. Adaptive optics based on machine learning: a review. *Opto-Electronic Advances* **5**, 200082 (2022).
15. Lohani, S. & Glasser, R. T. Turbulence correction with artificial neural networks. *Opt. Lett.* **43**, 2611-2614 (2018).
16. Li, J. et al. Turbulence compensation with a deep-learning-based method for orbital angular momentum-multiplexed free-space optical communication. *Nat. Commun.* **11**, 6149 (2020).
17. Lavery, M. P. J. Vortex instability in turbulent free-space propagation. *New J. Phys.* **20**, 043023 (2018).
18. Chen, M. & Lavery, M. *Optical angular momentum interaction with turbulent and scattering media in Structured Light for Optical Communication* 237-258 (Elsevier, 2021).
19. Torner, L., Torres, J. P. & Carrasco, S. Digital spiral imaging. *Opt. Express* **13**, 873 (2005).
20. Lavery, M. P. J. et al. Measurement of the light orbital angular momentum spectrum using an optical geometric transformation. *J. Opt.* **13**, 064006 (2011).
21. Cortes, C. & Vapnik, V. Support-vector networks. *Machine Learning* **20**, 273-297 (1995).
22. Mountrakis, G., Im, J. & Ogole, C. Support vector machines in remote sensing: A review. *ISPRS J. Photogramm. Remote Sens.* **66**, 247-259 (2011).
23. Sun, X. et al. High-precision recognition of OAM modes in a turbulent atmosphere via a support vector machine. *Opt. Express* **27**, 17939-17952 (2019).

24. Mendoza-Hernández et al. Laguerre-Gauss beams versus Bessel beams showdown: peer comparison. *Opt. Lett.* **40**, 3739-3742 (2015).
25. Durnin, J., Miceli, J. J. Jr & Eberly, J. H. Diffraction-free beams. *Phys. Rev. Lett.* **58**, 1499 (1987).
26. Gori, F., Guattari, G. & Padovani, C. Bessel-Gauss beams. *Opt. Commun.* **64**, 491–495 (1987).
27. Fried, D. L. Optical Resolution Through a Randomly Inhomogeneous Medium for Very Long and Very Short Exposures. *J. Opt. Soc. Am.* **56**, 1372-1379 (1966).
28. Dong, K. et al. Characterizing propagation and vortex-splitting dynamics of Bessel-Gaussian beams in short-range atmospheric conditions. *Opt. Express* **33**, 2878–2895 (2025).
29. Sokolova, M. & Lapalme, G. A systematic analysis of performance measures for classification tasks. *Inf. Process. Manage.* **45**, 427-437 (2009).
30. Peters, C., Cocotos, V. & Forbes, A. Structured light in atmospheric turbulence—a guide to its digital implementation: tutorial. *Adv. Opt. Photonics* **17**, 113-184 (2025).
31. Rodenburg, B. et al. Simulating propagation of orbital angular momentum beams in turbulence. *New J. Phys.* **16**, 063021 (2014).

## Acknowledgements

This work was supported by the 111 Project (B17035); the National Natural Science Foundation of China (62575227, U20B2059, 62231021, 61621005, 62201613); the Shanghai Aerospace Science and Technology Innovation Foundation (SAST-2022-069); and the Fundamental Research Funds for the Central Universities (ZYTS25121).

## Author contributions

Andrew Forbes and Lixin Guo jointly supervised this research. The initial concept was developed by Mingjian Cheng and Andrew Forbes. Wenjie Jiang developed the 2D OAM spectrum analysis method and processed the data using machine learning techniques. Lixin Guo provided support for the simulation implementation. Wenjie Jiang and Mingjian Cheng compiled and prepared the manuscript. All authors reviewed and approved the final manuscript.

## Competing interests

The authors declare no competing interests.